\documentclass[review]{elsarticle}

\usepackage[dvipdfmx]{graphicx}

\usepackage{lineno,hyperref}
\usepackage{tabularx}
\usepackage{color} 
\usepackage{numcompress}
\biboptions{sort&compress}
\newcolumntype{C}{>{\centering\arraybackslash}p{25mm}}
\newcolumntype{D}{>{\centering\arraybackslash}p{25mm}}
\modulolinenumbers[5]

\journal{CARBON}

\begin{document}

\begin{frontmatter}

\title{Tunable induced magnetic moment and in-plane conductance of graphene in Ni/graphene/Ni nano-spin-valve-like structure: a first principles study}

\author{Yusuf Wicaksono, Shingo Teranishi, Kazutaka Nishiguchi}
\author{Koichi Kusakabe\footnote{Corresponding author. Tel: +81 6-68506406. E-mail: kabe@mp.es.osaka-u.ac.jp (Koichi Kusakabe)}
}
\address{Graduate School of Engineering Science, Osaka University,

1-3 Machikaneyama-cho, Toyonaka, Osaka 560-8531, Japan}

\begin{abstract}
This study theoretically investigated the magnetic properties and electronic structure of a graphene-based nano-spin-valve-like structure. Magnetic nickel layers on both sides of the graphene were considered. A spin-polarized generalized-gradient approximation determined the electronic states. In an energetically stable stacking arrangement of graphene and two nickel layers, the anti-parallel spin configuration of the underlayer and overlayer magnetic moments had the lowest energy, which is in agreement with previous experimental studies. The spin density mapping and obtained band-structure results show that when the upper and lower Ni(111) slabs have an anti-parallel (parallel) magnetic-moment configuration, the carbon atoms of sublattices A and B will have an antiferromagnetic (ferromagnetic) spin configuration. A band gap at the Dirac cone was open when the alignment had an anti-parallel configuration and closed when the alignment had a parallel configuration. Therefore, the in-plane conductance of the graphene layer depends on the magnetic alignment of the two nickel slabs when the Fermi level is adjusted at the Dirac point. Both the magnetic properties and electronic structures of the Ni/graphene/Ni nanostructure cause the system to be a new prospective spintronic device showing controllable in-plane magnetoresistance.
\end{abstract}


\end{frontmatter}

\section{Introductions}

Graphene, a carbon-based 2D material with extraordinary in-plane charge mobility \citep{Geim20076183}, has attracted interest as an ideal material for application in microelectronics \citep{Novoselov2005438197,Morozov20081001016602} and sensing \citep{Schedin20076652}. Experimentally, graphene is synthesized by chemical vapor deposition (CVD) and yields well-ordered \textit{3m symmetry} on the substrate surface \citep{Shikin1999155,KKS2009457706}. Graphene grown on top of the transition metal enables the sensitive tuning of transport properties \citep{Abtew201355,Schultz20134494}. By contrast, graphene gives a unique magnetic response owing to edge states or by growing on top of the ferromagnetic material \citep{Abtew201355,Ziatdinov20148915,Ziatdinov20138711,Morishita2016858,Miyao2017863,Ziatdinov2017197,Weser2010961}. Its interesting tunable transport properties and unique magnetic response as well as its weak spin--orbit coupling and long spin scattering length \citep{Novoselov666} make graphene a prospective material for spintronics devices. 

Many studies have attempted to fabricate graphene-based spintronic devices as new potential materials for magnetic tunneling junction (MTJ) applications. The Ni(111) surface is the most commonly used metal contact to study graphene-based spintronics owing to its structure similar to that of graphene (the smallest lattice mismatch among other transition metals) and its strong hybridization with graphene \citep{Abtew201355,Dahal201462548}. Recent experimental and theoretical studies of graphene in a sandwich structure with Ni(111) suggested two structural models: (i)Ni/graphene/Ni contact, where graphene is used as a bridge between two nickel electrodes and the current flow is parallel to the plane, and (ii)Ni/Graphene/Ni spin valve structure, which consists of two Ni slabs sandwiching a graphene layer and the current flows out of the plane. A theoretical study of the Ni/graphene/Ni contact reported a high magnetoresistance ratio \citep{Cho2011115,Sato2012859,Liu2016111}. However, no experimental study has yet confirmed this. Meanwhile, the Ni/graphene/Ni spin-valve structure can be synthesized easily and shows the change of the magnetoresistance ratio due to an external magnetic field \citep{Mandal20124986,Iqbal2018205}. 
Although further satisfactory results are expected, a new strategy to overcome the low resistance and absence of bandgap in graphene is necessary to realize a spintronics device based on graphene.

This paper presents a perspective of conductivity change that differs from those of previous studies on Ni/graphene/Ni nanostructures: an in-plane conductivity change of graphene in a sandwich structure of monolayer graphene with Ni(111) slabs is proposed. In this study, instead of investigating the conductivity change perpendicular to the plane of the layered structure, the in-plane conductivity change of the graphene layer due to the magnetic alignment, i.e. anti-parallel and parallel configuration, between the upper and lower Ni slabs was considered. The controllability of the spin-dependent Dirac states in this hybrid structure was studied through the magnetic properties and electronic structure of graphene. 
To explore the magnetic properties of Ni/graphene/Ni nano-spin-valve-like structure, especially the induced magnetic moment properties of graphene due to the anti-parallel or parallel magnetic configuration of the upper and lower Ni slabs, a density functional theory (DFT) spin generalized gradient approximation (spin-GGA) calculation was done to determine the spin-dependent charge density and magnetic moment of a Ni/graphene/Ni nanostructure. A band structure calculation was done to show the indication of graphene in-plane conductivity change by observing the change in the Dirac cone characteristic. 

\section{Methodology}

\begin{figure}[tb]
	\centering
	\includegraphics[scale=0.45,trim={0cm 0cm 0cm 0cm},clip]{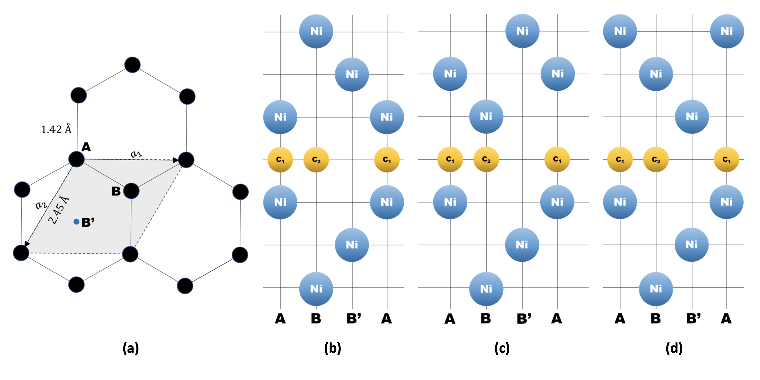}
	\caption{Model crystal structures of Ni/graphene/Ni systems. 
In the left panel (a), the top view of the graphene is depicted. 
While the proposed model crystal structures of the 3-layer Ni/graphene/3-layer Ni are shown 
in the (b) AA-stacking structure, (c) BA-stacking structure, and (d) B'A-stacking structure, where 
the stacking sequences from the graphene to the upper Ni slab are different from one another.}
	\label{fig:model crystal structure}
\end{figure}

The DFT-spin-GGA calculations were done using the Quantum ESPRESSO package \citep{qespresso}. Previous study reported that the original Perdew--Burke--Ernzerhof (PBE)\citep{PBE} functional determines wrongly the interlayer distance between the graphene and the nickel layer at the interfaces in the most stable structure of Ni/graphene system \citep{PBEincosistency1}. Therefore, the revised Perdew--Burke--Ernzerhof functional, called the PBEsol \citep{PBEsol} functional within the generalized gradient approximation (GGA), was used due to its consistency with experimental results in determining the interlayer distance. Ultrasoft pseudopotentials \citep{ultrasoft} were used to describe the electron--ion interaction. A kinetic energy cutoff of 50 Ry was used for wavefunctions to obtain a good convergence calculation. Since an appropriate k-point grid determined the convergence of the total energy calculation in this system, a k-point grid of 24$\times$24$\times$1 was used for all calculations. 

The adsorption of the graphene layer on the top of Ni(111) has four possible  structural stacking arrangements: (i) top/fcc-stacking--carbon atoms of graphene are placed on top of the first and third layers of Ni(111), (ii) top/hcp stacking--carbon atoms of graphene are placed on top of the first and second layers of Ni(111), (iii) hollow-stacking--carbon atoms of graphene are placed on top of the second and third layers of Ni(111), while the hollow site of graphene is placed on top of the first layer of Ni(111), (iv) bridge-stacking--bridge-site between two carbon atoms is placed on top of the first layer of Ni(111) \citep{Dahal201462548,Lahiri201113025001}. 

The consistency of the PBEsol functional in determining the most stable stacking arrangement and the interlayer distance between graphene and Ni(111) slab was verified by calculating the total energy of the Ni/graphene system using the four different stacking arrangements and compare the interlayer distance of the lowest total energy structure with previous experimental and theoretical studies. The result demonstrated that the top/fcc-stacking was the lowest energy state in comparison with the top/hcp-, hollow-, and bridge-stacking by 25.9, 122.0, and 226.4 meV, respectively. Meanwhile, the interlayer distance between graphene and Ni(111) slab of top/fcc-stacking was $2.08$\AA~. This result is in agreement with previous experimental and theoretical studies which reported top/fcc-stacking as the most stable structure with interlayer distance $\sim$ 0.21 nm \citep{PBEincosistency1,Dahal201462548,Lahiri201113025001,Kozlov20121167360}. However, although the top/fcc-stacking is the most preferable structure, the difference between top/fcc and top/hcp is small enough to allow both configurations to occur. 

Furthermore, the necessity of using van der Walls interaction in the calculation was verified by calculating the interlayer distance of the most stable structure, i.e., top/fcc-stacking, and comparing it with the calculation done without the inclusion of the van der Walls interaction. It was found that the interlayer distance difference was only $0.02$\AA, where the interlayer distances with the van der Walls interaction was $2.10$\AA. Although the interlayer distance with the inclusion of the van der Walls interaction was closest to the previous experimental studies, the small interlayer distance difference implies that the bonding between the nickel slab and graphene mainly comes from pd hybridization. Therefore, the contribution of the van der Walls interaction is omitted in this paper, and the main focus is placed on the covalent bond that comes from the pd hybridization of nickel atoms with carbon atoms.

The precise stacking arrangement of the Ni/graphene/Ni system has not yet been fully understood; the stacking of the Ni overlayer (slab) above the graphene must be verified. 
Since a relevant atomic configuration should usually be energetically lower than the others, this paper proposes three different arrangements of structural stacking for the upper part of the Ni/graphene interface, while top/fcc-stacking is determined for the lower part of the Ni/graphene by considering the most stable structural arrangement as shown in Fig. \ref{fig:model crystal structure}. The three different stacking arrangements are (i) AA-stacking, for which the first (third) layer of the upper Ni slab is placed on top of the A-site (B-site) carbon atoms of graphene, (ii) B'A-stacking, for which the first (second) layer of the upper Ni slab is placed on top of the B-site (A-site) carbon atoms of graphene, and (iii) BA-stacking, for which the second (third) layer of the upper Ni slab is placed on top of the B-site (A-site) carbon atoms of graphene. 

In the model structure, a vacuum space of at least 30\AA~ was inserted to avoid spurious interactions between the slab replicas. At first, the case of a three-layer nickel sandwich graphene layer was investigated. Then, the thickness of the nickel layers was varied from 1 to 6. To understand the magnetic configuration of the system, two initial magnetic configurations were set between the two nickel slabs: (i) anti-parallel configuration, for which the total magnetic moment of the upper and lower nickel slabs have an anti-parallel alignment, and (ii) parallel configuration, for which the total magnetic moment of the upper and lower nickel slabs have a parallel alignment. The total energies of the in-plane and out-of-plane magnetization directions were compared by adopting the noncollinear magnetism calculation with spin--orbit interaction terms. The in-plane magnetization direction was found to have an energy lower by only 1 meV. However, since this paper considers the functionality of in-plane conductance switching of graphene depending on the external magnetic field, the out-of-plane magnetization direction was chosen so that the anti-parallel and parallel of the Ni slabs could be controlled easily.

\section{Results and Discussions}

\subsection{\textit{Total energy and magnetic properties}}

\begin{table}
	\centering
	\caption{Total energy difference for various stacking arrangements of 3Ni/graphene/3Ni. The total energy is expressed in meV relative to the ground state (highlighted in bold font). APC(PC) refers to the antiparallel (parallel) magnetic alignments between the upper and lower nickel slabs }
	\begin{tabular}{|c|c|c|c|}
		\hline 
		\textbf{Magnetic} & \multicolumn{3}{c|}{\textbf{Total Energy (meV)}} 
		\\
		\cline{2-4}
		\textbf{Configuration} & \textbf{AA-stacking} & \textbf{B'A-stacking} & \textbf{BA-stacking} \\
		\hline
		APC & 70.1 & 156.4 & \textbf{0.0} \\
		\hline
		PC & 75.6 & 154.0 & 10.5 \\
		\hline
	\end{tabular}
	\label{tabel:totE-3NiGr3Ni}
\end{table}

The total energies of the Ni/graphene/Ni system for each of the stacking arrangements were considered in order to understand which of the three different stacking arrangements was the most energetically stable. Table \ref{tabel:totE-3NiGr3Ni} shows the total energy difference of 3-layer Ni/graphene/3-layer Ni relative to the lowest energy state for AA-stacking, B'A-stacking, and BA-stacking in the anti-parallel and parallel configurations. Among the three different stacking arrangements, B'A-stacking has the highest energy, both in the anti-parallel and parallel spin configurations. 
The highest energy state comes from the weakened bonds of the graphene layer with an upper Ni(111) slab. The weakened bond can be verified by observing the interlayer distance of the B'A stacking structure shown in Table \ref{tabel:interlayer_distance-3NiGr3Ni}. 

Previous experimental and theoretical studies of graphene growth on top of Ni(111) slabs suggest that the bonding occurs between an A-site carbon atom and the nickel atom closest to it. 
Indeed, the interlayer distance from the graphene to the nearest Ni(111) slab was concluded to be $\sim$ 0.21 nm \citep{Dahal201462548,Kozlov20121167360,Lahiri201113025001}. 
In the case of the 3-layer Ni/graphene/3-layer Ni nanostructure in the B'A-stacking arrangement, the distance between the graphene layer and upper Ni(111) slabs becomes 3.28\AA~ (or 3.26\AA) for the anti-parallel (parallel) configuration, although the interlayer distance between the graphene layer and Ni(111) underlayer is 2.07 \AA. This fact also indicates an anti-bonding characteristic because of the considerable increase in the interlayer distance from the typical distance of approximately 2.1\AA.

Furthermore, by comparing the bonding characteristics of the graphene and upper Ni(111) slabs in both the AA-stacking and BA-stacking structures with the anti-bonding nature of the B'A-stacking structure, it can be inferred that the construction of pd hybridization and chemical bonding between the carbon atoms in the graphene layer with the nickel atoms at the interface mainly come from the hybridization of the $d_{z^2}$ orbital of nickel atoms with the $p_z$ orbital of carbon atoms. Therefore, stacking the nickel layer on top of either sublattice A or sublattice B of carbon atoms in the graphene produces a stable Ni/graphene/Ni nanostructure. 

Comparing the stable structures of the 3-layer Ni/graphene/3-layer Ni, it was found that BA-stacking has a slightly smaller interlayer distance than AA-stacking, indicating that BA-stacking has a stronger pd hybridization than AA-stacking. This difference shows that BA-stacking has the lowest energy among the three stacking arrangements. The strong pd hybridization between a carbon atom of the graphene layer with the upper or lower Ni(111) layer at the interface of BA-stacking leads to a charge transfer from the nickel layers to the graphene.

 \begin{table}
	\centering
	\caption{Interlayer distance between the Ni(111) slabs with graphene for three different stacking arrangements. APC(PC) refers to the antiparallel (parallel) magnetic alignment between the upper and lower nickel slabs }
	\begin{tabular}{|c|c|c|c|c|c|c|}
		\hline 
		& \multicolumn{6}{c|}{\textbf{Interlayer distance (\AA)}} 
		\\
		\cline{2-7}
		\textbf{Layer} & \multicolumn{2}{c|}{\textbf{AA-stacking}} &  \multicolumn{2}{c|}{\textbf{B'A-stacking}} & \multicolumn{2}{c|}{\textbf{BA-stacking}} \\
		\cline{2-7}
		& \textbf{APC} & \textbf{PC} & \textbf{APC} & \textbf{PC} &\textbf{APC} & \textbf{PC}  \\
		\hline
		Graphene-Upper Ni slab & 2.18 & 2.20 & 3.28 & 3.26 & 2.07 & 2.07 \\
		\hline
		Graphene-Lower Ni slab & 2.18 & 2.20 & 2.07 & 2.07 & 2.07 & 2.07 \\ 
		\hline
	\end{tabular}
	\label{tabel:interlayer_distance-3NiGr3Ni}
\end{table}
Further investigation on the most stable structure, BA-stacking, was done by comparing two possible structure arrangements of the second and third layer namely, B(top-hcp)A-stacking--the second (third) layer of upper Ni(111) slab on top of hollow- (B-) site of graphene (the initial proposed structure of BA-stacking) and B(top-fcc)A-stacking--the second (third) layer of upper Ni(111) slab on the top of B- (hollow-) site of graphene. The result showed that B(top/hcp)A-stacking has higher energy by 13.5 meV compare to B(top/fcc) stacking. Although B(top/fcc) stacking has lower energy than B(top/hcp)A-stacking, the difference was considerably small enough to allow both configurations to occur. It implied that  the small different might not change the physical properties of the whole system.

\begin{figure}[tb]
	\centering
	\includegraphics[scale=0.4,trim={0cm 0cm 0cm 0cm},clip]{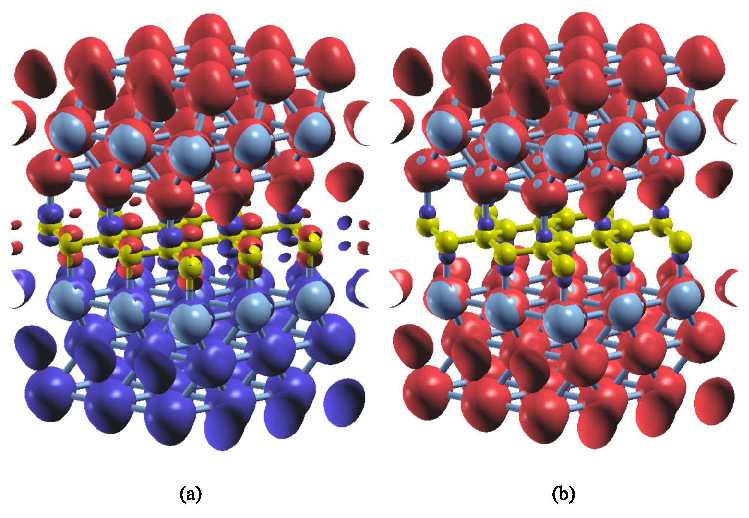}
	\caption{Spin-charge density of the 3-layer Ni/graphene/3-layer Ni in the (a) anti-parallel configuration (isovalue=0.00170) and (b) parallel configuration (isovalue=0.00168). The charge density in red represents spin-up electron density and that in blue represents spin-down electron density.  }
     \label{fig:3NiGr3Ni_spin-charge_density}
\end{figure}

Within the BA-stacking structure, the total energy of the anti-parallel configuration was lower than that of the parallel configuration by 10.5 meV. The anti-parallel configuration, which has a more stable energy state than the parallel configuration, is in agreement with experimental data reported by Mandal et al. \citep{Mandal20124986}. This total energy difference mainly comes from the magnetic configuration within the graphene layer.  

Spin--charge density mapping was created to understand the spin orientation of each atom in the 3-layer Ni/graphene/3-layer Ni. In Fig.\ref{fig:3NiGr3Ni_spin-charge_density}, the AFM configuration of the carbon atoms within the graphene layer, when the magnetic configuration between the upper and lower nickel slabs was anti-parallel, is shown. When the magnetic configuration between the upper and lower nickel slabs is parallel, a ferromagnetic (FM) configuration, which has a higher energy than AFM configuration, occurs within the graphene layer. AFM configuration is realized as the most stable state, except when an external magnetic field stabilizes FM configuration.  

The different magnetic configurations of the carbon atoms in the sublattices A and B within the graphene layer comes from the magnetic interaction between the Ni layer and graphene. Carbon atoms in graphene tend to have an anti-parallel spin configuration with the neighboring nickel atoms. This rule is the most relevant rule for the energetics of Ni/graphene/Ni structures. As a result of the spin anti-parallel, {\it i.e.,} locally anti-ferromagnetic, alignment between each pair of Ni and carbon atoms, AFM (FM) configurations in the graphene plane happens for the anti-parallel (parallel) configuration of Ni slabs. The local AFM configuration between a carbon atom and nickel atom will be considered in the subsequent sections by analyzing the hybridization nature of electron orbitals.

In the present model, the upper Ni(111) slab was fixed in a spin-up configuration, whereas the lower slab could be controlled to have either a spin-up or spin-down configuration. The nickel atoms on the upper Ni/graphene interface hybridized with carbon atoms on sublattice B, and since the upper Ni(111) slab was fixed to have a spin-up alignment, the carbon atoms on the sublattice B of graphene had the spin-down configuration. By contrast, the lower Ni(111) slab was hybridized with carbon atoms in the sublattice A. Since the lower Ni(111) slab can be controlled between the spin-up and spin-down configurations, the induced magnetic moment of the carbon atoms in sublattice A can also be controlled. 

\begin{table}
	\centering
	\caption{Magnetic moment for each atom of 3Ni/graphene/3Ni. AFM(FM) refers to the antiparallel (parallel) magnetic alignment between the upper and lower nickel slabs. }
	\begin{tabular}{|c|c|c|}
		\hline 
		\textbf{Atom} & \multicolumn{2}{c|}{\textbf{Magnetic Moment ($\mu_B$)}} 
		\\
		\cline{2-3}
		& \textbf{AFM configuration} & \textbf{FM configuration} \\
		\hline
		 C (sub-lattice A) & 0.0126 & -0.0048   \\
		 \hline
		 C (sub-lattice B) & -0.0128 & -0.0047 \\
		 \hline
		 Ni(top)-first layer & 0.2898 & 0.2645 \\
		 \hline
		 Ni(top)-second layer & 0.6150 & 0.5958 \\
		 \hline
		 Ni(top)-third layer & 0.6785 & 0.6732 \\
		 \hline
		 Ni(bottom)-first layer & -0.2770 & 0.2815 \\
		 \hline
		 Ni(bottom)-second layer & -0.6213 & 0.6050\\
		 \hline
		 Ni(bottom)-third layer & -0.6763 & 0.6670 \\
		 \hline
	\end{tabular}
	\label{tabel:magmom_3Ni-Gr-3Ni}
\end{table}

The carbon atoms of sublattices A and B in graphene tend to have an antiferromagnetic (AFM) configuration due to the half-filled $p_z$ orbital and Pauli's exclusion principle. This rule is often found in organic molecules in sp$^2$ hybridization or magnetic alternant hydro-carbon systems \citep{Itoh1967235,Wasserman196789}. The spin configuration of AFM is often considered as a realization of the spin-density-wave. Due to the nature of graphene, the carbon atoms of another sublattice will have the opposite alignment with the carbon atoms hybridizing with the nickel atoms. Therefore, the lowest energy state was found to be the anti-parallel configuration. 

Interestingly, in the parallel configuration, FM spin alignment appears within the graphene layer. However, the appearance of the FM configuration within graphene, where the spin moment has the opposite direction to the Ni layers, was unexpected. In the case of the Ni/graphene interface, i.e., graphene on a Ni substrate,  carbon atoms are placed in two different sites, A and B, where one of the sites were hybridized with Ni layer and the another not. The magnetic moment of carbon atoms that were not hybridized with Ni layer, which has a parallel spin to the nickel slab, was higher than the hybridized one, where the antiferromagnetic configuration in graphene appears as a whole. It implied that the Ni(111) slab and unhybridized carbon atoms had a stronger magnetic interaction. Oppositely, the FM configuration within graphene of Ni/graphene/Ni system implied another conclusion that the induced magnetic moment at the hybridized carbon atoms had a stronger magnetic interaction with the Ni(111) slabs than the unhybridized carbon atom. It was because the direction of the magnetic moment was reversed in the unhybridized carbon atoms. Therefore, a stronger magnetic interaction and the anti-parallel configuration between the hybridized carbon atom and the Ni layer were concluded. 

The magnetic configuration, which is initiated by the magnetic interaction of carbon and nickel atoms at the interface, within the graphene layer leads to the tunable magnetic moment configuration between the carbon atoms in sublattices A and B of graphene. The tunable induced magnetic moment between the AFM and FM configurations provides a new insight into the electronic structure of graphene, since the previous studies have only reported that the induced magnetic moment of graphene is either in the AFM configuration \citep{graphene-antiferro} or FM configuration \citep{graphene-ferro}. An additional remarkable point is that the total energy difference between the anti-parallel and parallel configurations is in the order of meV, which implies that the magnetic configuration change does not require an extremely high external magnetic field. 

\begin{table}
	\centering
	\caption{ Total energy differences between various numbers of the nickel layer in the BA-stacking arrangement. The total energy is expressed in meV relative to the ground state (highlighted in bold font). APC(PC) refers to the antiparallel (parallel) magnetic alignment between the upper and lower nickel slabs.}
	\begin{tabular}{|c|C|D|}
		\hline 
		\textbf{Number of Nickel Layers} & \multicolumn{2}{c|}{\textbf{Total Energy (meV)}} 
		\\
		\cline{2-3}
		& \textbf{APC} & \textbf{PC} \\
		\hline
		 1 & \multicolumn{2}{c|}{total energy same (non-magnetic)}   \\
		 \hline
		 2 & \textbf{0.0} & 9.9 \\
		 \hline
		 3 & \textbf{0.0} & 10.4 \\
		 \hline
		 4 & \textbf{0.0} & 1.3 \\
		 \hline
		 5 & \textbf{0.0} & 7.6 \\
		 \hline
		 6 & \textbf{0.0} & 6.1 \\
		\hline
	\end{tabular}
	\label{tabel:totE-xNiGrxNi}
\end{table}

The total energy was calculated for various thicknesses to show the consistency of the most stable magnetic configuration and the orientation of the Ni/graphene/Ni nanostructure's magnetic moment. Table \ref{tabel:totE-xNiGrxNi} shows the total energy for all variations of the number of Ni layers from 1 to 6 in the anti-parallel and parallel configurations. For the 1-layer Ni structure at both sides of the graphene, the lowest state is a non-magnetic state. The variation of the number of Ni layers from 2 to 6 implies that the anti-parallel configuration has the lowest energy, although the total energy difference between the anti-parallel and parallel configurations is different for each number of layers. The charge-spin density mapping also shows the magnetic configuration of Ni/graphene/Ni. 

\subsection{\textit{Analysis of electron orbitals for Ni/graphene/Ni with 1-layer Ni slabs}}

As shown in the last section, in the case of 1-layer Ni growth on both sides of graphene, 
the Ni/graphene/ Ni nanostructure becomes non-magnetic. 
The non-magnetic property of the 1-layer Ni/graphene/1-layer Ni occurred due to the charge transfer from the $d_{z^2}$ orbital of Ni atoms within 1-layer Ni to the $p_z$ orbital of carbon atoms, either for sublattices A or B within graphene, which fully filled the $p_z$ orbitals of the carbon atoms and caused the net magnetic moment to become zero. 

\begin{figure}[tb]
	\centering
	\includegraphics[scale=0.35,trim={0.5cm 0cm 0cm 0cm},clip]{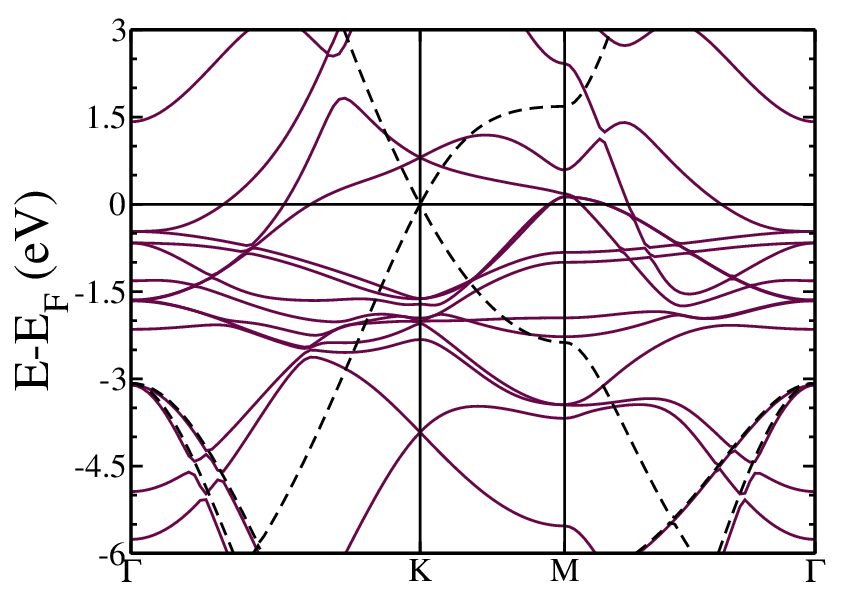}
	\caption{Electronic band structure of 1-layer Ni/graphene/1-layer Ni in a non-magnetic state. The black dashed line is the band structure of pristine graphene, to be used as reference. The Dirac cone characteristic can be seen around the K-point high-symmetry and near the Fermi energy as well as around -4 eV.}
	\label{fig:1NiGr1Ni_bandstructure}
\end{figure}

The electronic band structure of 1-layer Ni/graphene/1-layer Ni is shown in Fig.~\ref{fig:1NiGr1Ni_bandstructure}. This band structure implies relevant orbital natures in Ni/graphene/Ni structures. 
Here, every band around the Fermi level was characterized. 

Eight bands from Ni 3$d$ branches can be seen just below the Fermi level at $E_F=0$, i.e., at the $\Gamma$ point.   
Other two bands of 3$d$ go down to approximately -5eV or below. 
This energy is even below the top of the sp$^2$-$\sigma$ bands appearing at -3 eV. 
At the $\Gamma$ point, 
the $\pi$ band of graphene, which has an energy of around -7.5eV in the pristine graphene, 
is rather stabilized. This is attributed to the pd hybridization. 
Since the flat graphene becomes deformed with staggered modification, 
the orbital nature of $\pi$ becomes a little closer to that of the sp$^3$-hybridized orbital. 
This is the secondary reason why such an apparent drop is observed in the $\pi$ band. 
Roughly speaking, the eight 3$d$ bands of Ni run nearly horizontally across the whole Brillouin zone. 
Meanwhile, the remaining two branches coming from the $d_{z^2}$ bands suffer strongly 
from the pd hybridization. 

Local Wannier orbitals of $p_z$ and $d_{z^2}$ symmetries were defined at each carbon site and nickel atom, respectively. These orbitals are represented by 
$\phi_{p_z,\eta}({\bf r}-{\bf R}_i)$ and $\varphi_{z^2,\zeta}({\bf r}-{\bf R}_i)$. 
Here, the orbitals are defined for each unit cell indexed by $i$. The vector ${\bf R}_i$ 
represents the position of the $i$-th unit cell. 
The index $\eta$ denotes the sublattice site A or B of graphene. 
$\zeta$ denotes the sublattice sites for the upper and lower Ni layers. 
However, by limiting the argument to the BA-stacking, $\zeta$ is identified as $\eta$, 
since every Ni atom is just on (or below) a carbon atom at the A (or B) site. 
Therefore, it is enough to consider simplified notations $\phi_{i,p_z,\eta}$ and $\varphi_{i,z^2,\eta}$. 

For graphene, the Dirac points appear at the special points of K and K'. 
Since the three-fold rotational symmetry, glide-reflection symmetry, and 
chiral symmetry of the original graphene is preserved 
for the AB stacking of the 1-layer Ni/graphene/1-layer Ni, 
the Dirac point is maintained in this hybrid structure. 
More precisely, Dirac cones were doubled in the band structure of Fig.~\ref{fig:1NiGr1Ni_bandstructure}. 
At the K point, around $E-E_F\sim$ -4eV, there are lower Dirac cones. 
There are also more upper Dirac cones above the Fermi level.  
The lower (upper) cone consists of the bonding (anti-bonding) level of 
$\phi_{i,p_z,\eta}$ and $\varphi_{i,z^2,\eta}$. 

The orbital energy of $p_z$ is much lower than that of nickel $d_{z^2}$. Therefore, 
the charge transfer is from nickel to carbon. The magnetic moment of carbon is not 
easily induced in this sense. 
The anti-bonding nature of the upper Dirac cone close to the Fermi level 
is important for the induced magnetism of graphene attached to nickel slubs. 

Here, it should be noted that the effective potential at each sublattice is the same as the effective potential for the non-magnetic solution as far as the Ni /graphene/Ni system maintains 
the structural symmetry. Topologically, the arrangement of the effective Wannier orbital 
$\tilde\phi_{i,\eta,\pm}$ for both the upper and lower Dirac cones is the honeycomb lattice. 
Therefore, no gap opening happens for two Dirac cones in Fig.~\ref{fig:1NiGr1Ni_bandstructure}. 

When spin--density distribution causes a staggered contribution in 
a spin-dependent effective potential, a gap opening phenomenon occurs. 
This point is relevant for the discussion in the next section. 

\subsection{\textit{Effect of magnetic configuration on the electronic structure and in-plane conductance}}

The appearance of the induced magnetic moment on graphene and 
strong pd hybridization between the nickel and carbon atoms on graphene 
cause the Dirac cone characteristics of the graphene band structure to change. 

The opening gap of the Dirac cone has been reported by previous theoretical studies of the Ni/graphene interface, 
where the gap size between the spin majority and spin minority channels is 
different due to the induced magnetic moment difference between sublattices A 
(carbon atoms with pd hybridization with nickel atoms) and B. 
The magnetic moment of carbon atoms can be also expressed as the spin--charge density 
difference between spin-up and spin-down ($n_\uparrow ({\bf r})-n_\downarrow ({\bf r})$). 
The charge density difference between spin-up and spin-down is proportional to 
the Stoner gap formed on the energy level difference between spin-up and spin-down. 
In the Ni/graphene interface case, the magnetic moment of carbon atoms on sublattice A 
is smaller than that of sublattice B; the Stoner gap in the carbon atoms on sublattice A 
is also smaller compare to that of sublattice B. 

A relevant question here is whether not the Ni/graphene/Ni structures show any qualitatively different features from a single Ni/graphene interface.  
. 

\begin{figure}[tb]
	\centering
	\includegraphics[scale=0.45,trim={0.5cm 0cm 0cm 0cm},clip]{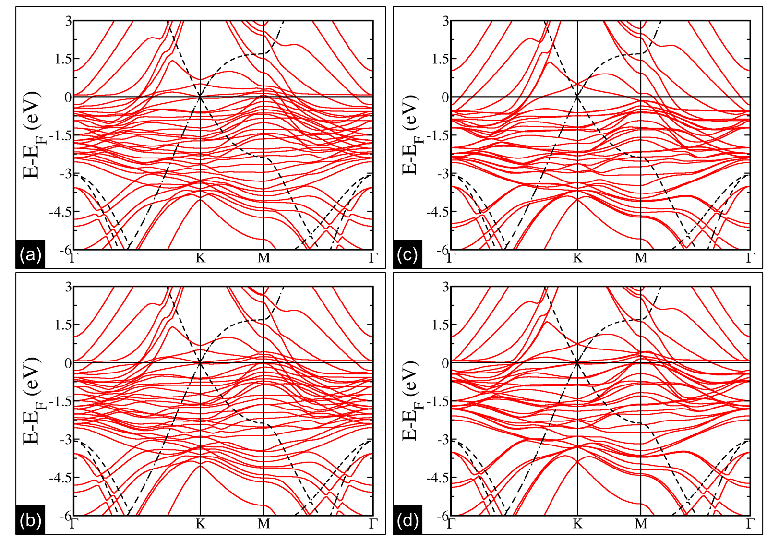}
	\caption{3-layer Ni/graphene/3-layer Ni band structure in the anti-parallel configuration for the (a) spin majority channel and (b) spin minority channel, and in parallel configuration for the (c) spin majority channel and (d) spin minority channel. The black dashed line is the band structure of the pristine graphene to be used as reference. The Dirac cone characteristic can be seen around the K-point high-symmetry and near the Fermi energy.}
	\label{fig:3NiGr3Ni_bandstructure}
\end{figure}

Figure \ref{fig:3NiGr3Ni_bandstructure} shows the band structure of 
3-Ni layer /graphene/3-Ni layer spin-valve-like structure. 
The strong hybridization between the carbon $p_z$ orbital with nickel $d_{z^2}$ on the top and bottom is shown as the creation of bonding and anti-bonding Dirac cones and several level anti-crossing points 
in the band structure. 
Although the band structure of the carbon p$_z$ orbital changes quite significantly, 
the characteristics of Dirac cones on high symmetry K-points 
can still be recognized for both the anti-parallel and parallel configurations. 
Here, two Dirac cones are seen near the Fermi energy and around -4eV below the Fermi energy. 

For the case of anti-parallel configuration, Table \ref{tabel:magmom_3Ni-Gr-3Ni} shows that 
the induced magnetic moment of the carbon atoms on sublattice A is in the opposite configuration 
to that on sublattice B. 
The magnetic moment of carbon atoms on sublattice A has a spin-up configuration, 
which means that the spin-up charge density 
is higher than the spin-down charge density ($n_{A\downarrow}$), 
while carbon atoms on sublattice B have the opposite configuration. 

An integrated spin-charge density $n_{\eta,\uparrow}$, and $n_{\eta,\downarrow}$ was introduced, where $\eta$ is either A or B. 
The value of $n_{\eta,\sigma}$ is assumed to be obtained by integrating 
$n_\sigma({\bf r})$ within an ionic radius of the carbon atom, where 
$\sigma=\uparrow$ or $\downarrow$. 
Table \ref{tabel:magmom_3Ni-Gr-3Ni} also shows that 
the magnetic moment between the carbon atoms 
on sublattices A and B are comparably equal 
in the anti-parallel configuration. 
Therefore, 
\[ n_{A,\downarrow} = n_{B,\uparrow} > n_{A,\uparrow} = n_{B,\downarrow}.\]

The spin-GGA effective potential has a contribution approximately proportional to 
the spin moment, or the spin density, $n_{\eta,\uparrow}-n_{\eta,\downarrow}$. 
This spin-dependent potential is lowered for the majority spin when the local 
spin density is given by the majority, while the minority spin feels higher energy. 
In other words, the energy of the up-electron decreases when the up-spin density increases, 
but increases when the down-spin density increases around the given point. 

Therefore, the up-spin and down-spin suffer from different effective potentials $v_{{\rm eff},\sigma}({\bf r})$ in the anti-parallel configuration. 
Taking a spin-dependent Wannier function, $ \tilde\phi_{i,\eta,-,\sigma}({\bf r}) $, 
created similar to that in the last section, a simple rule can be derived for an averaged 
spin-dependent contribution for the potential 
$\langle \tilde\phi_{i,\eta,-,\sigma}| \bar{v}_{{\rm eff},\sigma} | \tilde\phi_{i,\eta,-,\sigma}\rangle$ 
with respect to the Wannier basis. 
\begin{eqnarray}
\langle \tilde\phi_{i,A,-,\uparrow}| \bar{v}_{{\rm eff},\uparrow} | \tilde\phi_{i,A,-,\uparrow}\rangle
&>&
\langle \tilde\phi_{i,B,-,\uparrow}| \bar{v}_{{\rm eff},\uparrow} | \tilde\phi_{i,B,-,\uparrow}\rangle, 
\label{1steq}
\\
\langle \tilde\phi_{i,A,-,\downarrow}| \bar{v}_{{\rm eff},\downarrow} | \tilde\phi_{i,A,-,\downarrow}\rangle
&<&
\langle \tilde\phi_{i,B,-,\downarrow}| \bar{v}_{{\rm eff},\downarrow} | \tilde\phi_{i,B,-,\downarrow}\rangle, 
\label{2ndeq}
\\
\langle \tilde\phi_{i,A,-,\uparrow}| \bar{v}_{{\rm eff},\uparrow} | \tilde\phi_{i,A,-,\uparrow}\rangle
&=&
\langle \tilde\phi_{i,B,-,\downarrow}| \bar{v}_{{\rm eff},\downarrow} | \tilde\phi_{i,B,-,\downarrow}\rangle, 
\label{3rdeq}
\\
\langle \tilde\phi_{i,B,-,\uparrow}| \bar{v}_{{\rm eff},\uparrow} | \tilde\phi_{i,B,-,\uparrow}\rangle, 
&=&
\langle \tilde\phi_{i,A,-,\downarrow}| \bar{v}_{{\rm eff},\downarrow} | \tilde\phi_{i,A,-,\downarrow}\rangle.
\label{4theq}
\end{eqnarray}
Eqs.~(\ref{1steq}) and (\ref{2ndeq}) conclude that a gap opens at the Dirac point 
due to the broken chiral symmetry. 
From Eqs.~(\ref{3rdeq}) and (\ref{4theq}), it can be said that the gap size is the same 
for both the up and down spins. 
Therefore, the Stoner gap size formed on the spin majority and minority channels 
are equal, as shown  in Fig.~\ref{fig:3NiGr3Ni_bandstructure}(a). 

Consider a realistic device by taking the attachment of pristine graphene as leads of the Ni/graphene/Ni structure. From this perspective, the momentum of electron in the graphene layer can be conserved by considering the specular reflection effect on the edge of the Ni layer. Looking closely at the K-point near the Dirac cone, separation of Dirac bands from the Ni S-band in the first Brillouin zone happens. Assume that the edge of the Ni layer perpendicular to the current path is periodic, sharply cut with no impurities, and the specular reflection of electron on graphene occurs. If the wavefunction of electron is larger than the atomic distance between the nickel on the edge perpendicular to the current path, the momentum of the Dirac band electrons will conserve. 

In contrast, Fig.~\ref{fig:3NiGr3Ni_bandstructure}(b) shows the band structure of 
3-layer Ni/graphene/ 3-layer Ni in the parallel configuration, where it satisfied an equality and inequality as follows
\begin{eqnarray}
\langle \tilde\phi_{i,A,-,\uparrow}| \bar{v}_{{\rm eff},\uparrow} | \tilde\phi_{i,A,-,\uparrow}\rangle
&=&
\langle \tilde\phi_{i,B,-,\uparrow}| \bar{v}_{{\rm eff},\uparrow} | \tilde\phi_{i,B,-,\uparrow}\rangle, 
\label{5theq}
\\
\langle \tilde\phi_{i,A,-,\downarrow}| \bar{v}_{{\rm eff},\downarrow} | \tilde\phi_{i,A,-,\downarrow}\rangle
&=&
\langle \tilde\phi_{i,B,-,\downarrow}| \bar{v}_{{\rm eff},\downarrow} | \tilde\phi_{i,B,-,\downarrow}\rangle, 
\label{6theq}
\\
\langle \tilde\phi_{i,A,-,\uparrow}| \bar{v}_{{\rm eff},\uparrow} | \tilde\phi_{i,A,-,\uparrow}\rangle
&>&
\langle \tilde\phi_{i,A,-,\downarrow}| \bar{v}_{{\rm eff},\downarrow} | \tilde\phi_{i,A,-,\downarrow}\rangle, 
\label{7theq}
\end{eqnarray}
Although the Dirac cone still survives, the energy level between the Dirac cone on 
the spin majority channel is lower than the Dirac cone on the spin minority channel 
because of the characteristics of the ferromagnetic material. 
Note that the above discussion does not directly reveal the relations between the orbital energies of 
the Wannier basis. This is because the orbital energy is determined by the whole 
character of $v_{{\rm eff},}({\bf r})$,  where the majority spin gains energy by 
the large spin moment at the Ni atoms. 

The opened gap of the Dirac electrons at the high symmetry K-point in the spin-majority and -minority channels for the anti-parallel spin configuration implies an increase of resistance on the in-plane conductance of the graphene layer compared to the pristine graphene. In contrast, the survival of the Dirac cone for the parallel magnetic configuration of the Ni(111) layers implies that
the in-plane conductance via the K-point contribution will give lower 
resistivity compared to that in the anti-parallel spin configuration.

Since the anti-parallel configuration is the lowest state, we can start 
from the high resistance state having the opened gap at the Dirac cone.
Using an external magnetic field, the spin moment direction can be reversed.
In the parallel configuration, the gap is closed in both majority and minority channels. Thus, the structure can act as a field-induced switch for the electron current.

In case of unbalanced spin moments below and above the 
Graphene layer, the applied field may reverse one of the spin directions, maintaining the other stronger spin moment undisturbed.
After switching off the field, the system can return to an energetically 
stabler anti-parallel spin configuration than the parallel configuration.
Alternatively, even when a residual spin moment remains for the zero fields, the weakly locked moment can be reversed by applying a magnetic field in the opposite direction. This controllability is essential to consider an application of this system.

\begin{figure} [tb]
	\centering
	\includegraphics[scale=0.28,trim={0cm 0cm -0.5cm 0cm},clip]{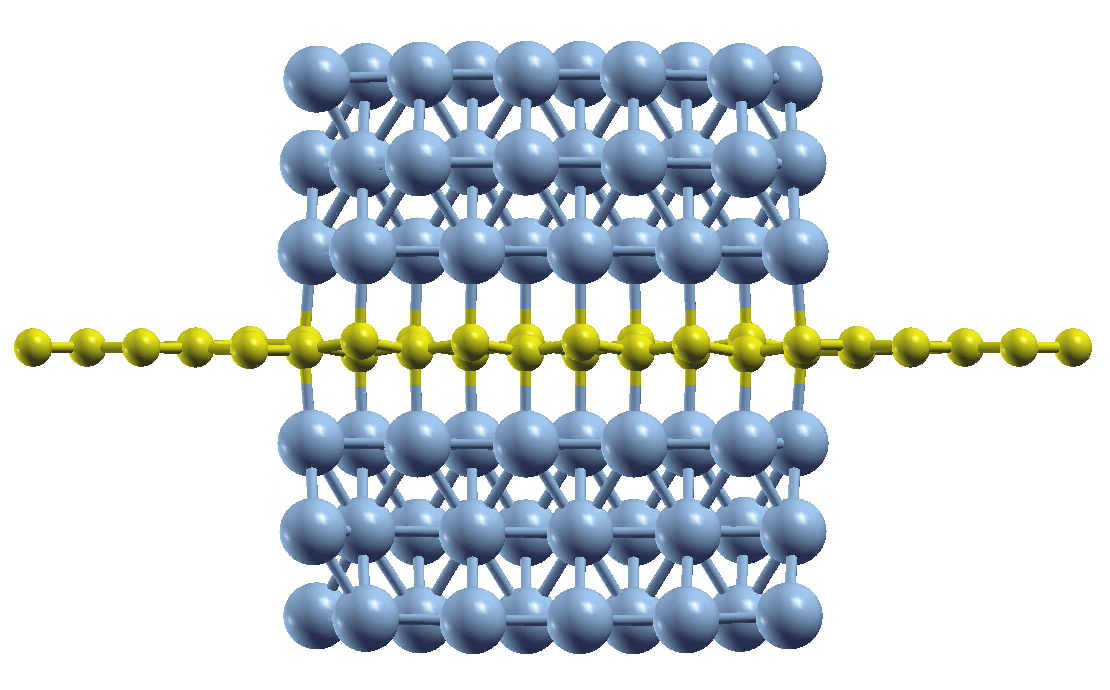}
	\caption{Scattering region of Ni/Graphene/Ni nano-spin-valve-like structure created on the part of a graphene plane.}
	\label{fig:junction}
\end{figure}

As an example, a junction system is proposed here. Figure~\ref{fig:junction} shows a Ni/Graphene/Ni nano-spin-valve-like structure created on the part of a graphene plane. The local Ni structures above and below graphene are optimized by DFT simulations. This magnetic region gives scattering of electron waves. Pristine graphene parts at both sides are used as the electrodes. Through the one-dimensional device structure, we consider a current path from the left to right electrodes for both spins.

As shown in Fig.~\ref{fig:junction}, the edges of the Ni(111) layers are atomically sharp. In case of a long Ni part in the perpendicular direction for the current path, the specular reflection/transmission for electron waves can be obtained. The Dirac cone characteristic is dependent on the Ni spin configuration. Therefore, change in the in-plane conductance along the graphene layer across the scattering region is expected.

The Fermi energy (the chemical potential) can be slightly shifted from the charge neutrality point. Note that, in the anti-parallel spin configuration, net spin current is not observed even when the current path is via the Ni(111) layers; in contrast, in the parallel configuration, the spin-majority and -minority channels created by Ni structures have different scattering processes. In the graphene layer, a closed gap phenomenon is observed, which allows the spin majority channel for the spin alignment to be less resistive than in the anti-parallel spin configuration. In addition, owing to the spin blockade effect for current paths via the Ni(111) layers, a stronger conduction is observed for a selective spin direction only. Thus, we can easily create a spin-current switching.

At low temperatures, when coherent transmission becomes dominant for the current conduction, where a lateral element of the momentum along the edge is conserved, the spin filtering phenomenon is expected to be enhanced. Consider a wider Ni(111) layer region along the one-dimensional conduction path. Dirac electrons tunnel via states in the window opened in the Brillouin zone around the K-point. In the anti-parallel configuration, when only coherent tunneling paths along the structure determine the transmission probability, the current conduction is effectively blocked. This is because the gap is open and the Dirac electrons are blocked. Since transmission probability depends on the electronic structure at the scattering region, we can conclude a spin-configuration dependent tunneling phenomenon in this junction structure.
This effect will lead to a spin filtering effect of electron current in the graphene electrode. In the anti-parallel configuration, this junction system also leads to a spin Hall effect \citep{spin-hall} owing to the spin--orbit interaction originating in the Ni layers.

\begin{figure} [tb]
	\centering
	\includegraphics[scale=0.37,trim={0cm 0cm -0.5cm 0cm},clip]{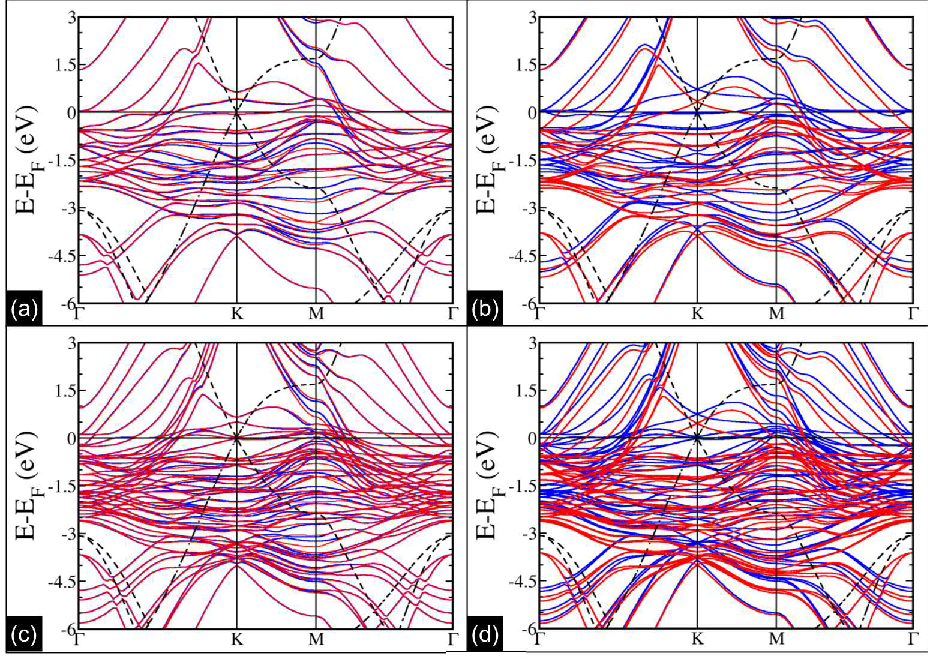}
	\caption{Bandstructure of 2-layer Ni/graphene/2-layer Ni in the (a) anti-parallel configuration and (b) parallel configuration; 4-layer Ni/graphene/4-layer Ni in the (c) anti-parallel configuration and (d) parallel configuration; and (e) 1-layer Ni/graphene/1-layer Ni. In (a), (b), (c), and (d), red represents the spin-majority channel, whereas blue represents the spin-minority channel. The 1Ni-layer/graphene/1-Ni layer bandstructure shows unpolarized bandstructure because the system becomes a non-magnetic material. The black dashed line is the bandstructure of the pristine graphene, taken as a reference}
	\label{fig:xNiGrxNi_bandstructure}
\end{figure}

The opened gap structure at the Ni(111)/graphene/Ni(111) interface suggests a moment-direction-dependent characteristic for the lateral conductance with a perpendicular current path. In this conventional use for a Ni-based spin valve, the gap opening effect should be considered properly. The determination of the transmission probability is future work for the present study.

The opening gap of the Dirac cone in the anti-parallel configuration and the survival of the Dirac cone in the parallel configuration are consistent for 2-layer and 4-layer nickel, as shown in figure 
\ref{fig:xNiGrxNi_bandstructure}. For the number of Ni layers more than four, the band-structure becomes extremely complex, making recognizing the Dirac cone characteristics difficult. However, owing to a preserved symmetry, we can expect the same physics.

\section{Conclusions}

In this study, DFT (spin-GGA) calculation on 3-layer Ni/graphene/3-layer Ni shows that among the three stacking arrangements, the BA-stacking is the stablest stacking structure, and within BA-stacking, the anti-parallel configuration of upper and lower Ni(111) slabs has lower energy than the parallel configuration. This finding is in agreement with the previous experimental study. The two magnetic arrangements of Ni(111) slabs affect the configuration of the induced magnetic moment of graphene. 

In case of the anti-parallel configuration between Ni(111) slabs, carbon atom sublattices A and B of graphene have an AFM configuration. By contrast, in case of the parallel configuration between two Ni(111) slabs, the configuration of sublattices A and B of graphene have an FM configuration. The AFM configuration of graphene's carbon atoms leads to the gap opening on the Dirac cone of the graphene bandstructure; meanwhile, the FM configuration bandstructure shows the survival of the Dirac cone both for spin-majority and -minority channels. 

The opening and survival of the Dirac cone will affect the in-plane conductance of graphene in which the open gap Dirac cone configuration will have a higher resistance than the close gap Dirac cone configuration. Further conductivity calculation must be conducted to understand how much does the opening and closing of the gap due to induced magnetic moment on graphene's carbon atoms affect the conductivity of the graphene layer. 

\section*{Acknowledgment}
The calculations were done at the computer centers of Kyushu University
and ISSP, University of Tokyo. 
This work is partly supported by JSPS KAKENHI Grant No.
JP26400357, JP16H00914 in Science of Atomic Layers, and JP18K03456.
Y. W. gratefully acknowledges scholarship support from the Japan International Cooperation Agency (JICA) within the "Innovative Asia" Program, ID Number D1707483.
The authors thank S. K. Saha for fruitful discussions.

\bibliography{mybibfile}

\end{document}